\documentclass[prd,nofootinbib,tightenlines,showkeys,showpacs]{revtex4}
\usepackage{amsfonts}
\usepackage{amssymb}
\usepackage{textcomp}
\everymath{\displaystyle}
\usepackage[dvips]{graphicx}
\usepackage{longtable}
\usepackage{subfigure}
\usepackage{indentfirst}

\usepackage[latin1]{inputenc}
\def\be{\begin{equation}}
\def\ee{\end{equation}}
\def\beq{\begin{eqnarray}}
\def\eeq{\end{eqnarray}}
\def\bn{\begin{eqnarray*}}
\def\en{\end{eqnarray*}}

\def\w{\omega}

\def\a{\alpha}
\def\b{\beta}

\def\d{\delta}

\def\t{\theta}
\def\T{\Theta}

\def\pd{\partial}
\def\e{\epsilon}

\def\m{\mu}

\def\t{\theta}

\def\bra{{\langle}}
\def\ket{{\rangle}}
\def\bp{{\bf p}}

\def\bk{\bf k}

\def\cH{{\cal{H}}}
\def\cP{{\cal{P}}}

\def\cA{{\cal{A}}}

\def\cS{{\cal{S}}}
\def\cF{{\cal{F}}}

\begin{document}

\title{{\small\hfill IMSC/2010/06/08}\\
{\small\hfill SU-4252-908}\\
Thermofield dynamics and twisted Poincar\'e symmetry on Moyal space-time}
\author{A. P. Balachandran\footnote{bal@phy.syr.edu}}
\affiliation{Physics Department, Syracuse University
Syracuse, NY, 13244-1130, USA }
\author{T. R. Govindarajan\footnote{trg@imsc.res.in}}
\affiliation{The Institute of Mathematical Sciences
C. I. T. Campus Taramani, Chennai 600 113,India}



\begin{abstract}
On Moyal space-time, one can implement twisted Poincar\'e symmetry with the resultant
modification of symmetrization and anti-symmetrization postulates for bosons and fermions.
We develop the thermofield approach of
Umezawa and Takahashi on such a spacetime preserving the twisted Poincar\'e symmetry of the
underlying quantum field theory(qft).
Implications of this twisted Poincar\'e symmetry for qft's at finite temperature are
pointed out.
\end{abstract}

\maketitle

\section{ Introduction}
Quantum fields on noncommutative spacetimes like the Moyal spacetime have been studied in 
recent times extensively\cite{wroclaw1,wroclaw2}. 
So also are there studies on fuzzy geometries like coadjoint orbits of 
compact Lie groups\cite{balbook}. These studies indicate novel modifications of symmetrisation and 
antisymmetrisation postulates conventionally used for bosons and fermions. They also 
establish new phases which are novel\cite{sondhi,denjoe,trg}. The existence 
of such features makes it interesting
to study quantum field theories  (qfts) on the Moyal spacetime at finite 
temperature.  A study along these line was initiated in \cite{akofor} 
using linear response theory. Here we will instead examine
quantum fields at finite temperature on the  noncommutative Moyal 
space-time using Umezawa and Takahashi's thermofield dynamics approach\cite{umezawa}.
Thus this work is complemented by \cite{akofor} and also by \cite{sachin}.

The thermofield  approach uses ``mirror" fields and Bogoliubov transformations 
to construct the thermal ``vacuum" state. Once such a state is constructed the 
various expectation values of operators in this state give the thermal distributions
effectively. 

In Sec. II we will briefly review thermofield dynamics for commutative space-time. 
In Sec. III we will extend the above to the Moyal space-time. In Sec IV, we present 
an analysis of interacting field theories in the above approach. 

\section{Review of Thermofield dynamics}
In an important contribution, Umezawa and Takahashi \cite{umezawa} 
constructed a quantum field theory known as thermofield dynamics (TFD),
in which the thermal average of an observable is given by the vacuum expectation value 
in an extended Hilbert space. This is achieved by constructing a thermal state
$\mid 0(\beta) \ket$ such that 
\be
\bra A \ket ~\equiv ~ \frac{Tr e^{-\b H} A}{Tr e^{-\b H}} ~=~ \bra 0(\beta)\mid A 
\mid 0(\beta) \ket.
\label{thermal}
\ee
where $\b ~=~\frac{1}{kT},~k$ is the Boltzman constant and $T$ is the temperature.

For this one starts with a Hamiltonian $H$ acting on a  Hilbert space $\cH$ and following Umezawa
and Takahashi extend the latter  to $\cH \otimes \tilde{\cH}$. Here $\tilde{\cH}$
is also a Hilbert space isomorphic to $\cH$. The total Hamiltonian $\hat{H}$
is given by 
\be
\hat{H} ~=~ H~ - \tilde{H}
\ee
where we will give the details of $\tilde{H}$ below.

If $P_\mu$ is the four-momentum for $\cH$ with $P_0~=~H$, then $\tilde{P}_\mu$
on $\cH \otimes \tilde{\cH}$ has a similar form (see below) 
and $\hat{P}_\mu ~=~ P_\mu - \tilde{P}_\mu$.
In the above by $H$ and $\tilde{H}$ ($P_\mu$ and $\tilde{P}_\mu$) we mean 
$H\otimes I$ and $I\otimes \tilde{H}$ ($P_\mu \otimes I$ and 
$I\otimes \tilde{P}_\mu$) while $\tilde{P}_0~=~\tilde{H}$.

In TFD, the degrees of freedom are doubled. To every observable operator $O$,
we assign a tilde conjugate operator $\tilde{O}$. These conjugations obey the following
rules:
\be 
(O_1O_2)\tilde{} ~=~ \tilde{O}_1 \tilde{O}_2,~ 
(cO)\tilde{} ~=~ c^* \tilde{O},~ 
(O^\dagger)\tilde{} = (\tilde{O})^\dagger,~
(O\tilde{})\tilde{} ~=~ \pm O~(+~for~boson,-~for~fermion)
\ee
The untwisted free field $\phi(x) \left(\tilde{\phi}(x)\right)$ of mass $m$ has the conventional 
mode expansion:
\be
\phi(x)~\left(\tilde{\phi}(x)\right)=~
\frac{1}{(2\pi)^3}\int \frac{d^3k}{2\w(k)}
\left(c_{\bk} (\tilde{c}_{\bk})e^{-i{k.x}}~+~
c^\dagger_{\bk}(\tilde{c}_{\bk}^\dagger)e^{ik.x}\right), ~~~\w_k~=~+~\sqrt{{\bk}^2~+~m^2}.
\ee
In the above $c_{\bk}, c_{\bk}^\dagger$ and $\tilde{c}_{\bk}, \tilde{c}_{\bk}^\dagger$ are 
the standard Fock 
space annihilation and creation operators of $\cH, \tilde{\cH}$. 
\be 
\left[c_{\bk},~c_{{\bk}'}^\dagger\right]~=~\left[\tilde{c}_{\bk},\tilde{c}_{{\bk}'}^\dagger\right]~=~
(2\pi)^3~(2\w(k))\d^3({\bk}~-~{\bk '}).
\ee
Then $\mid 0(\b)\ket$
is  the ``thermal vacuum 
state" of annihilation and creation operators $\a_{\bk},~\a_{\bk}^\dagger$ and 
$\tilde{\a}_{\bk}, \tilde{\a}_{\bk}^\dagger$. They are obtained by a  
Bogoliubov transformation of the creation (annihilation) operators 
$c_{\bk} (c_{\bk}^\dagger)$ as follows:
\be
\a_{\bk} ~=~ e^{-i G} c_{\bk} e^{i G};~~ \tilde{\a}_{\bk} ~=~ e^{-i G} \tilde{c}_{\bk} e^{i G}
\label{Udef}
\ee
where 
\be
G~=~ -~\frac{i}{(2\pi)^3}~\int \frac{d^3k}{2\w(k)} \T ({\bk}) (c_{\bk}\tilde{c}_{\bk}~-
~c_{\bk}^\dagger \tilde{c}_{\bk}^\dagger)
\label{Gdef}
\ee
Here 
$$
\tanh^2\T(k)~=~ e^{-\b \w (k)}, 
$$ 
for free fields. 

The exact expression for $\a_{\bk},\tilde{\a}_{\bk}$ are:
\bn
\a_{\bk}~&=&~\cosh\T(k) c_{\bk}~-\sinh \T(k) \tilde{c}_{\bk}^\dagger \\
\tilde{\a}_{\bk}~&=&~\cosh\T(k) \tilde{c}_{\bk}~-~\sinh \T(k) c_{\bk}^\dagger
\en
One can define Bogoliubov transformed field $\phi_D(x)\left(\tilde{\phi}_D(x)\right)$:
\be
\phi_D(x)\left(\tilde{\phi}_D(x)\right)~
=~\frac{1}{(2\pi)^3}\int \frac{d^3k}{2\w(k)}
\left(\a_{\bk}~(\tilde{\a}_{\bk})~e^{-ik.x}~+~
\a^\dagger_{\bk}~(\tilde{\a}^\dagger_{\bk})e^{ik.x}\right).
\label{phiD}
\ee
The thermal ``vacuum'' state condition is that $\mid 0(\b)\ket $ is 
annihilated by $\a_{\bk},\tilde{\a}_{\bk}$:
\be
\a_{\bk}~\mid 0(\b) \ket ~=~ \tilde{\a}_{\bk}~\mid 0(\b)\ket ~=~ 0.
\ee
It is given by
\beq
\mid 0(\b) \ket ~=~U(\T)|0,\tilde{0}\ket ~&=&~ \int \frac{d^3k}{2\w(k)}\sum_{n_k}
\frac{1}{\sqrt{Z_{\bk}(\b)}}e^{-\frac{\b E_n(\w_k)}{2}}\mid n_{\bk},\tilde{n}_{\bk} \ket,\\
E_n(\w(k))~&=&~n\w(k),\qquad U(\T)~=~e^{-iG}. \nonumber
\eeq
In the above $Z_{\bk}(\b)~=~Tr_{n_{\bk}}e^{-\b H}$ (with trace over all states with fixed $n_{\bk}$) 
is the normalisation 
factor and $E_n(\w_k)$ is the energy of
a state with $n_{\bk}$ particles all of momentum ${\bk}$. 
$\mid n_{\bk},\tilde{n}_{\bk}\ket$ is the state with each 
of $n_{\bk}$ particles, all of momenta ${\bf k}$, in
$\cH$ and $\tilde{\cH}$. (Thus $E_1(\w(k))~=~\w(k)$.) 
With this definition of the thermal vacuum, it is obvious that Eq(\ref{thermal}) follows.

The operator $N_{\bk} ~=~ c_{\bk}^\dagger c_{\bk} ~-~\tilde{c}_{\bk}^\dagger \tilde{c}_{\bk}$
commutes with $G$.  
Hence:
\be
N_{\bk} \mid 0(\b)\ket ~=~ \bra 0(\b) \mid N_{\bk} ~=~0.
\ee

Given the above it is easy to work out the Green's functions for the free
theory at finite temperature. For this purpose consider the two fields $\phi, \tilde{\phi}$
as a column $\Phi$:
\be
\Phi(x)~=~ \pmatrix{\phi(x) \cr \tilde{\phi}(x)}
\ee
Then 
\beq 
i~G_0(x,y)~&=&~ \bra 0 \tilde{0}|T(\Phi(x)\Phi^T(y)|0 \tilde{0}\ket \label{greenfn} \\
~&=&~ 
\pmatrix
{\bra 0 \tilde{0}|T(\phi(x)\phi(y))|0 \tilde{0}\ket   & 0\cr
              0 & \bra 0 \tilde{0}|T(\tilde{\phi}(x)\tilde{\phi}(y))|0 \tilde{0}\ket}\nonumber\\
~&=&~\frac{1}{(2\pi)^4}\int d^4k ~{(iG_0(k))}~ e^{-ik.(x-y)}
\eeq
where 
\be
G_0(k) ~=~ \pmatrix{\frac{1}{k^2-m^2+i\e} & 0 \cr
                      0 & \frac{1}{k^2-m^2-i\e} }
\ee
On the other hand the thermal Green's function is given by:
\beq
i~G_\b (x,y)~&=&~ \bra 0,\b|T(\Phi(x)\Phi(y)|0,\b\ket \\
~&=&~\bra 0 \tilde{0}|T(U(\T)^\dagger \Phi(x)U(\T)U(\T)^\dagger\Phi(y)U(\T)|0 \tilde{0}\ket\\
~&=&~\frac{1}{(2\pi)^4}\int d^4k~ {(iG_\b(k))}~ e^{ik.(x-y)}
\eeq
where $U(\T) = e^{-iG}$ is  defined as in Eqs.(\ref{Udef}) and 
\be 
G_\b(\bk)~=~\pmatrix{\frac{1}{k^2-m^2+i\e} -2\pi i \d(k^2-m^2) N(\bk) & 2\pi i \d(k^2-m^2) N(\bk)\cr
                   2\pi i \d(k^2-m^2) N(\bk) & \frac{1}{k^2-m^2-i\e} - 2\pi i \d(k^2-m^2) N(\bk)}
\ee
where $N({\bk}) ~=~ \sinh^2 \T(k)$.
 
Now one can diagonalise the above thermal Green's function  by a suitable linear combination
of $\phi$ and $\tilde{\phi}$ fields. For this purpose one starts from Eq(\ref{greenfn}) and 
inserts $U(\T)^\dagger U(\T) $ as follows:
\beq
i~G_0(x,y)~&=&~ \bra 0 \tilde{0}|T(U(\T)^\dagger U(\T)\Phi(x)\Phi^T(y)U(\T)^\dagger U(\T))
|0 \tilde{0}\ket \nonumber \\
&=&~\bra 0(\b)|T(\Xi_D(x)\Xi^T_D(y))|0(\b) \ket 
\eeq 
where we have defined 
$$
\Xi_D(x)~=~U(\T)\Phi(x)U(\T)^{\dagger} ~=~\pmatrix{\chi_D(x) \cr \tilde{\chi}_D(x)}
$$
which agrees with Eq(\ref{phiD}). ($U(\T)$ conjugates each entry in the column of $\Phi(x)$.)
In the next section, we will see how this formalism  extends 
to the Moyal plane and 
what are the changes that can be expected. 
\section{Thermofield dynamics on the Moyal plane}
In the Moyal plane $\cA_\t(R^4)$ described by the commutator,
\be
\left[ X_\m, X_\nu \right] ~=~ i\t_{\m\nu},
\ee
the ideas of TFD can be incorporated. 

We will first point out some important features 
about the Moyal plane and the action of the symmetry transformations on it.

The multiplication map $m_\t$ in the algebra $\cA_\t(R^4)$ is defined by
the $*$ product, 
\be
m_\t(f~\otimes~g)(x)~=~f(x)*g(x)~=~\left(f~e^{i\T^{\mu\nu}\overleftarrow{\pd}_\m \overrightarrow{\pd}_\nu}
g\right)(x)
\ee
where $f,g \in \cA_\t(R^4)$. Further the Poincar\'e group action on tensor prducts 
$\cA_\t(R^4)\otimes \cA_\t(R^4)\otimes \cdots \otimes \cA_\t(R^4)$ cannot 
be implemented in the conventional way. 
What is required is a twisted action 
of its group algebra on these tensor products. 
This leads to important changes in the algebra of creation/annihilation
operators and statistics. 
The outcome are the relations:
\be 
a_p~a_q~=~a_q~a_p~e^{ip\wedge q},~a_p~a_q^\dagger~=~e^{-ip\wedge q}~a_q^\dagger~a_p~+~2p_0~
\d^3(p-q)
\ee
where $A\wedge B~=~A_\mu\t^{\mu\nu}B_\nu$. For details see Balachandran et al . \cite{bal}. 

The operators $a_p,a_p^\dagger$ can be obtained from the standard annihilation and creation operators
$c_p,c_p^\dagger$ through the following ``dressing transformation'':
\be
a_p~=~c_p~e^{-\frac{i}{2}p\wedge P}~~~~a_p^\dagger~=~c_p^\dagger~e^{\frac{i}{2}p\wedge P}
\label{dress}
\ee

The twisted quantum field $\phi_\t$ can also be obtained by a dressing transformation 
from the untwisted quantum field:
\be
\phi_\t(x) ~=~ \phi_0(x) ~e^{\frac{1}{2}\overleftarrow{\pd}\wedge P}, \qquad \phi_0(x) := \phi(x)
\label{fielddress}
\ee
Here $P_\mu$ is the total momentum operator.

We now discuss how these `dressing transformations' generalise to thermal field
theories. We must first construct the twisted analogues 
$\a_{\bk}, \a^\dagger_{\bk}, \tilde{\a}_{\bk}, \tilde{\a}^\dagger_{\bk}$ 
of the creation and the annihilation operators. 
For this purpose, we need to generalise
Eqs.(\ref{Udef}) and (\ref{Gdef}). The important point to note in this context is that,
the total momentum operator in thermofield theory 
is the difference of momentum operators for $\phi$ and $\tilde{\phi}$ fields:
\be   
P_\mu (\phi_0,\tilde{\phi}_0) ~= P_\mu (\phi_0) - P_\mu (\tilde{\phi}_0)~~\equiv~P_\mu~-~\tilde{P}_\mu
\label{totalP}
\ee
This is required for the dressing transformation since $\a_{\bk}$ for example
involves the annihilation operators of $\phi$ and creation operators of $\tilde{\phi}$ fields.

The required Bogoliubov transformations for the twisted case are:
\beq 
\a^\T_{\bk} ~&=&~\left(\cosh\T(k)~c_{\bk}-~\sinh\T(k)\tilde{c}^\dagger_{\bk}\right)
e^{-\frac{i}{2}k\wedge P(\phi,
\tilde{\phi})}\\
\tilde{\a}^\T_k~&=&~\left(\cosh\T(k)~\tilde{c}_k~-~\sinh\T(k)c^\dagger_k\right)
e^{-\frac{i}{2}k\wedge P(\phi,\tilde{\phi})}
\label{twistbogoliubov}
\eeq
and their adjoints.

The twisted field at zero temperature is:
\beq
\phi_\t(x)~&=&~\frac{1}{(2\pi)^3}\int \frac{d^3p}{2p_0}
\left(a_{\bp} e^{-ip.x}~+~a^\dagger_{\bp}e^{ip.x}\right) \nonumber \\
\tilde{\phi}_\t(x)~&=&~\frac{1}{(2\pi)^3}\int \frac{d^3p}{2p_0}
\left(\tilde{a}_{\bp} e^{-ip.x}~+~\tilde{a}^\dagger_{\bp}e^{ip.x}\right).
\label{twist0}
\eeq
In the same way we can define the twisted `diagonal' fields at finite temperature:
\beq
\chi_{\t D}(x)~&=&~\frac{1}{(2\pi)^3}\int \frac{d^3p}{2p_0}
\left(\a_{\bp}^\T~e^{-ip.x}~+~\a^{\dagger\T}_{\bp}e^{ip.x}\right) \nonumber \\
\tilde{\chi}_{\t D}(x)~&=&~\frac{1}{(2\pi)^3}\int \frac{d^3p}{2p_0}
\left(\tilde{\a}_{\bp}^\T e^{-ip.x}~
+~\tilde{\a}^{\dagger\T}_{\bp}e^{ip.x}\right).
\eeq  
As the number operators commute with the twist, the thermal vacuum is independent of 
$\t_{\mu\nu}$. The twisted thermal vacuum $\mid 0(\b)\ket $ is defined by
\be
\a^\T_{\bk}| 0(\b) \ket ~=~ \tilde{\a}_{\bk}^\T~| 0(\b) \ket ~=~0
\label{tfvac}
\ee
For later use, we also define the twisted finite temperature fields $\phi_{\t,\b}$:
\beq
\phi_{\t,\b}~&=&~\frac{1}{(2\pi)^3}\int \frac{d^3p}{2p_0}
\left(a_{\bp,\b}~e^{-ip.x}~+~a_{\bp,\b}^\dagger e^{ip.x}\right)\\ 
&=&~\phi_0~e^{\frac{1}{2}\overleftarrow{\pd}\wedge P(\phi_0,\tilde{\phi}_0)},~~~~ 
a_{\bp,\b}~=~c_\bp~e^{-\frac{i}{2}p\wedge P(\phi_0,\tilde{\phi}_0)},~~
a_{\bp,\b}^\dagger~=~c_\bp^\dagger~e^{\frac{i}{2}p\wedge P(\phi_0,\tilde{\phi}_0)} \nonumber
\eeq
In the above in $\phi_{\t,\b}(x)$ the subscript $\b$ is added to 
emphasize that the twist involves $P_\mu(\phi_0,\tilde{\phi_0})$ (Eq. \ref{totalP}) 
which annihilates the thermal vacuum.

We also would like to remark  that
in our approach to gauge fields, the latter are {\it{not}} twisted and are associated with commutative
spacetime. So,  its  fields operators are not twisted.

\subsection{Wightman functions and propagators}
One can obtain the Wightman functions for the  ``free" theory:
\bn
\bra 0(\b)|(\phi_\t(x)\phi_\t(y)) | 0(\b)\ket~&=&~
\bra 0,0\mid (U^\dagger(\T)\phi_\t(x)\phi_\t(y)U(\T))
\mid 0,0\ket\\
 ~&=&~ i~W_\t^\b(x-y)~=~i~e^{-i\frac{1}{2}\pd_x\wedge\pd_y}~
W_0^\b(x-y)~=~i~W_0^\b(x-y).
\en
Hence we get $\Delta_{F,\t}^\b(x-y)$, the Feynman paropagator as:
\be
\Delta_{F,\t}^\b(x-y)~=~\bra 0(\b)|T(\phi_\t(x)\phi_\t(y)) | 0(\b)\ket~
=~\int \frac{d^4k}{(2\pi)^4} ~(i\Delta_{F,\t}^\b(k))~e^{-ik.(x-y)}
\ee
where 
\be
\Delta_{F,\t}^\b (k)~=~ \frac{1}{k^2-m^2+i\e}~-~2\pi i \d(k^2-m^2) N(k)\equiv \Delta_{F,{\t=0}}^\b (k)~
\ee
The above behaviour of 2-point function is independent of $\t$
because of translational invariance of the theory on the Moyal spacetime.
(We assume that $H$ and hence $\tilde{H}$ are spacetime translational invariant.)
Even though the two-point function is $\t$-independent, the higher order functions do depend 
on $\t$ and do not factorize
into sums of  products of two-point functions as it happens in zero temperature field theory.

The 4-point function already  reflects the important difference between  Moyal
and commutative space time. Consider
\bn
\cF(x_1,x_2,x_3,x_4)~&=&~_\t\bra 0(\b)\mid (\phi_\t(x_1)\phi_\t(x_2)\phi_\t(x_3)\phi_\t(x_4))
\mid0(\b)\ket_\t \\
~&=&~\bra 0,\tilde{0}\mid (U^\dagger(\T)\phi_\t(x_1)\phi_\t(x_2)\phi_\t(x_3)\phi_\t(x_4)U(\T))
\mid  0,\tilde{0}\ket \\
~&=&~\int \prod_i \frac{d^4p_i}{(2\pi)^4}~W_{\t}^\b(p_1,p_2,p_3,p_4)~e^{i\sum p_i\cdot x_i}
\en
Then 
\beq
W_{\t}^\b(p_1,p_2,p_3,p_4)~=&&W_{\t}^\b(p_1)W_{\t}^\b(p_3) \left[\d(p_1-p_2)\d(p_3-p_4)+~ 
\d(p_1-p_4)\d(p_2-p_3)\right] \nonumber \\
&&+~e^{ip_1\wedge p_2}W_{\t}^\b(p_1)W_{\t}^\b(p_2) \d(p_1-p_3)\d(p_2-p_4)\label{green}
\eeq
In the above $W_{\t}^\b(p)~=~\frac{1}{(2\pi)^4}\int e^{-ik.x}W_{\t}^\b(x)~
=~\frac{1}{(2\pi)^4}\int e^{-ik.x}W_{0}^\b(x)$
The higher order functions can also be worked out systematically.

Interestingly Eq.(\ref{green}) is the Bogoliubov transformed 4-point function of the twisted 
Poincar\'e invariant Wightman function given in an earliar zero temperature field 
theory\cite{bal}. 
 
{\it{Remark:}} The discussion till now has dealt with free or possibly interaction representation
fields. Remarks on Heisenberg fields will be made later.

\section{Scattering theory at finite temperature}
There are several equivalent approaches to scattering theory on commutative spacetimes, a few of
the well-known being the following: 

(1) The interaction representation scattering theory (2) The LSZ  formalism. 
(3) The Yang-Feldman appoach.

These approaches are not necessarily equivalent on noncommutative spacetimes even at zero temperature
\cite{DFR,Piacitelli}. Elsewhere \cite{iruv,baltrg} we have discussed the first two approaches 
on the Moyal spacetime at zero temperature in detail. We shall now generalise them to the finite 
temperature thermofield theory. 
\subsection{Scattering amplitudes in the interaction representation}
As a preliminary, we note that the free Hamiltonian is not affected by $\t_{\mu\nu}$ or $\b$,
it is the same as for $\t_{\mu\nu}~=~T~=~0$ for both matter and gauge fields. 

\noindent{\it{Without gauge fields:}}\\
We can consider the case of the real scalar field $\phi_0$ for $\t_{\mu\nu}~=~0, T~=~0$ 
as an illustration.
In the interaction representation, it is a free field. Consider an 
interaction Hamiltonian $H_I$
such as:
\be
H_I~=~\lambda \int d^3x ~:\phi_0^4: 
\label{intH}
\ee
If $\t_{\mu\nu}~\neq 0$, but $T~=~0$, it becomes $H^\t_I$ which for eq.(\ref{intH}) is 
\be
H^\t_I~=~\lambda \int d^3x ~:\phi_\t*\phi_\t*\phi_\t*\phi_\t:
\label{twistbeta}
\ee
where $\phi_\t$ is defined by eq.(\ref{twist0}) and normal ordering is 
with regard to $a_p,a_p^\dagger$ of 
eq.(\ref{dress}).

If both $\t_{\mu\nu}$ and $T$ are nonzero, then $H^\t_I$ becomes $H^{\t,\b}_I (H^{\t,\infty}~=~
H^\t_I)$  which for eq.(\ref{twistbeta}) is:
\be
H_I^{\t,\b}~=~\lambda \int d^3x ~:\phi_{\t,\b}*\phi_{\t,\b}*\phi_{\t,\b}*\phi_{\t,\b}: 
\ee
where the normal ordering is with respect to the $a^{\b}_{\bk},a^{\dagger \b}_{\bk}$
operators of eq.(\ref{twistbeta}). 

The interaction representation S-matrix $S_I^\t$ for $T~=~0$ is:
\be
S_I^\t~=~ T~\exp \left(-i \int dt  H^\t_I\right).
\ee
We showed in earliar papers that $S_I^\t$ is independent of $\t$:
\be
S_I^\t~=S_I^0~=S_I~=~T~\exp~\left(-i\int dt H_I\right).
\ee
The proof uses only properties of the field $\phi_0$, the *-product and translational 
invariance. Hence by the same arguments, the S-matrix $S_I^{\t,\b}$ is also independent of 
$\t_{\mu\nu}$ and $T$:
\be
S_I^{\t,\b}~=~T~\exp~\left(-i\int dt H_I^{\t,\b}\right)~=~S_I.   
\ee

The incoming and outgoing state-vectors are affected by $\t_{\mu\nu}$ and $\b$ 
since the Fock vacuum 
$~~\mid 0,0 \ket$ should be replaced by $\mid 0(\b) \ket$ and the 
particle states are created by repeated 
applications $a_{\bk}^{\b \dagger}$'s. 

Note that this is the correct procedure since (1) the free Hamiltonian is $\t_{\mu\nu}$ and $T$
independent so that the thermal vacuum has no dependence on $\t_{\mu\nu}$, and (2) we do 
not want to include the tilde excitations in the incoming and outgoing particles, so that particle
states which participate in scattering are to be created by $a_{\bk}^{\b \dagger}$'s 
from $\mid 0(\b)\ket$.

Consider a scattering process at temperature $T$ on the Moyal plane. If the incoming momenta are
$q_j,~~j~=~N+1,N+2, \cdots ,N+M$ and the outgoing momenta are $-q_k,~~k~=1,2, \cdots, N$, then 
incoming and outgoing state vectors at temperature $T$ are:
\beq 
\mid q_{N+M},q_{N+M-1}, \cdots, q_{N+1} \ket_{\t,\b}~&=&~a^\dagger_{q_{N+1},\b} a^\dagger_{q_{N+2},\b}
\cdots a^\dagger_{q_{N+M},\b}\mid 0(\b) \ket , \label{in}\\
\mid -q_{N},-q_{N-1}, \cdots, q_{1} \ket_{\t,\b}~&=&~a^\dagger_{-q_{N},\b} a^\dagger_{-q_{N-1},\b}
\cdots a^\dagger_{-q_{1},\b}\mid 0(\b) \ket .
\label{out}
\eeq
The scattering amplitude is the matrix element of $S_I$ between these vectors.

\noindent{\it{With gauge fields}:} 

In our approach, the gauge fields are not twisted. As a consequence, the twisted interaction Hamiltonian
$H_I^{\t,\b}$ splits into three parts :
\be
H_I^{\t,\b}~=~ H_I^{M,\t,\b}~+~H_I^G~+~ H_I^{M-G,\t,\b}.
\ee
Here $H_I^{M,\t,\b}~$ contains only matter fields and they are twisted, while $H_I^G~$ involves only
gauge fields and they are not twisted. The term $H_I^{M-G,\t,\b}$ involves both matter and gauge
fields with the former alone twisted. In each term, products of matter fields involve *,
but those of gauge fields with themselves or with matter fields require special
treatment. For details see \cite{baltrg}.

In the interaction representation, the S-Matrix $S^\t_I$ now depends on $\t_{\mu\nu}$:
\be
S_I^{\t,\b}~=~ T \exp \left(-i \int dt H_I^{\t,\b}(x) \right)
\ee
If there are incoming or outgoing gauge particles, we should include them in eqs.(\ref{in})
and (\ref{out}) by acting 
with their creation operators on $\mid 0(\b) \ket$. 
The scattering amplitude is the expectation value 
of $S_I^{\t,\b}$ between these extended vectors (\ref{in}) and (\ref{out}).
\subsection{The LSZ formalism}

\noindent{\it{Without gauge fields}}\\
We first consider the LSZ scattering amplitude for $\t_{\mu\nu}~=0$ for both $T~=~0$ and 
$T \neq 0$ and for spinless particles. We use the thermofield formalism. 

For $T = 0$, for ingoing and outgoing momenta as in eqs.(\ref{in}) and (\ref{out}), the scattering amplitude 
is 
\beq
S^0(-q_N,-q_{N-1},\cdots -q_1;&&q_{N+M},q_{N+M-1},\cdots ,q_{N+1}) \nonumber \\
=\int &&\prod^{N+M}_{i=1} d^4x_i~e^{-iq_i\cdot x_i}~i(\partial_i^2+m^2)
G_{N+M}^{0}(x_1,x_2,\cdots,x_{M+N}) 
\label{cLSZ}
\eeq
where $G_{N+M}^0$ is:
\beq 
G_{N+M}^0(x_1,x_2,\cdots,x_{M+N})) &=& T e^{\frac{i}{2}\sum_{I<J}\partial_{x_I}
\wedge \partial_{x_J}} W_{N+M}^{0} (x_1, \cdots x_{N+M}) \nonumber \\
:&=&~T~W_{N+M}^{0} (x_1, \cdots x_{N+M})
\label{GN0}
\eeq 
and $W_{N+M}^0$ are Wightman functions at $T ~=~0$:
\be
W_{N+M}^0~=~ ( 0 \mid \Phi_0(x_1)\cdots \Phi_0(x_{N+M})\mid 0 ) 
\ee
Here $\Phi_0$ are Heisenberg fields and $\mid0)$ is the vacuum state annihilated by 
$\cP_\mu$, the four momentum of the fully interacting theory :
\be
\cP_\mu \mid 0 )~=~0
\ee
In the twisted case, still at $T~=~0$ we have argued (see \cite{baltrg}) that $S^0$ is modified to
$S^\t$ where
\beq
S^\t(-q_N,-q_{N-1},\cdots -q_1;&&q_{N+M},q_{N+M-1},\cdots ,q_{N+1}) \nonumber \\
~=\int~&&\prod^{N+M}_{i=1} d^4x_i~e^{-iq_i\cdot x_i}~i(\partial_i^2+m^2) 
G_{N+M}^{\t}(x_1,x_2,\cdots,x_{M+N})~~~~
\label{ncLSZ0}
\eeq
where $G_{N+M}^{\theta}$ is:
\begin{eqnarray}
G_{N+M}^{\theta} (x_1, \cdots x_{N+M}) &=& T e^{\frac{i}{2}\sum_{I<J}\partial_{x_I}
\wedge \partial_{x_J}} W_{N+M}^{0} (x_1, \cdots x_{N+M}) \nonumber \\
:&=&~T~W_{N+M}^{\theta} (x_1, \cdots x_{N+M})
\label{GNtheta0}
\end{eqnarray}
and $W^\t_{N+M}$ are Wightman functions with $\t_{\mu\nu} \neq 0$.
\be
W^\t_{N+M} ~=~ ( 0 \mid \Phi_\t(x_1)\cdots \Phi_\t(x_{N+M}) \mid 0 ), ~~
\Phi_\t~=~\Phi_0~e^{\frac{1}{2}\overleftarrow{\pd}\wedge \cP}.
\label{twistwightman}
\ee
The twisted Heisenberg field $\Phi_\t$ generalises (\ref{fielddress}).

In arriving at eq.(\ref{twistwightman}), it is important to note that:
(a) $\mid 0 ) $ is stable under evolution. It is in fact $\t_{\mu\nu}$ independent
(b) as the energy-momentum operator $\cP_\mu$ of in- and out- fields $\cP_\mu^{in,out}$
are all equal to $\cP_\mu$, and $\Phi_0$ approaches in- and out- fields $\Phi_0^{in,out}$ as
$x_0 \rightarrow \mp \infty$, $\Phi_\t$ approaches their twisted versions: 
\beq
\Phi_\t & \stackrel{x_0 \rightarrow \mp \infty} \longrightarrow & \Phi_{\t}^{in,out} \\
\Phi_\t^{in,out}~&=&~ \Phi_0^{in,out}~e^{\frac{1}{2}\overleftarrow{\pd}\wedge \cP}, 
~~~~~\cP_\mu~=~\cP_\mu^{in,out}    
\eeq

We now generalise $S^\t$ to the finite temperature scattering amplitude $S^{\t,\b}$.
For this purpose we introduce the tilde Heisenberg fields $\tilde{\Phi}_0$ 
and their energy momentum operator $\tilde{\cP}_\mu$. Then if we can consistently replace
$\cP_\mu$ by $\cP_\mu ~-~\tilde{\cP}_\mu$ and define an ``exact" Heisenberg vacuum
$\mid 0(\b))$ and appropriate twisted Heisenberg field $\Phi_{\t,\b}$ we can write 
down $S^{\t,\b}$.

But we see that this is straightforward.  Thus

\noindent (a) $\mid 0(\b))$ can be constructed from the in-states of $\cP_\mu$ and 
$\tilde{\cP}_\mu$:
\beq
\mid 0(\b)) ~&=&~ \sum_N \int \frac{1}{\sqrt{Tr_N~e^{-\b P_0}}}~\prod_{i=1}^N d\mu(k_i) 
~e^{-\b P_0\over 2}\mid k_1,\cdots k_N \ket_{in}
\mid \tilde{k}_1,\cdots \tilde{k}_N \ket_{in} \\
d\mu (k_i)~&=& \frac{d^3k_i}{2|k_0|}.
\eeq
where $Tr_N$ denotes trace in the $N$-particle sector.
Since $\cP_\mu~=~\cP_\mu^{in}~=\cP_\mu^{out}$ and  $\tilde{\cP}_\mu~=~\tilde{\cP}_\mu^{in}~=~
\tilde{\cP}_\mu^{out}$ it is clear that:
\be
(\cP_\mu~-~\tilde{\cP}_\mu) \mid 0(\b))~=~0
\ee
It is also plausible that
\beq
\mid 0(\b)) ~&=&~ \sum_N \int ~\frac{1}{\sqrt{Tr_N~e^{-\b P_0}}}~
\prod_{i=1}^N d\mu(k_i) ~e^{-\b P_0\over 2}~\mid k_1,\cdots k_N \ket_{out}
\mid \tilde{k}_1,\cdots \tilde{k}_N \ket_{out} \\
d\mu(k_i)~&=& \frac{d^3k_i}{2|k_0|}.
\eeq

\noindent (b) The twisted Heisenberg field at temperature $T$ is:
\be
\Phi_{\t,\b}~=~\Phi_0 e^{\frac{1}{2}\overleftarrow{\pd}\wedge (\cP -\tilde{\cP})}
\ee
Since $\Phi_0$ fulfills asymptotic condition and $\cP_\mu~-~\tilde{\cP}_\mu ~=~ 
\cP_\mu^{in,out}~-~\tilde{\cP}_\mu^{in,out}$, $\Phi_{\t,\b}$ at least formally
obeys the asymptotic conditions
\be
\Phi_{\t,\b}(x) \stackrel{x_0 \rightarrow \mp \infty} \longrightarrow \Phi_0^{in,out}
e^{\frac{1}{2}\overleftarrow{\pd}\wedge ({\cP}_\mu^{in,out}-\tilde{\cP}_\mu^{in,out})}
\ee
as required.

The in- and out-states now are:
\beq
\mid q_{N+M},q_{N+M-1}, \cdots, q_{N+1})^{in}_{\t,\b}~&=&~a^{\dagger in}_{q_{N+1},\b} 
a^{\dagger in}_{q_{N+2},\b}
\cdots a^{\dagger in}_{q_{N+M},\b}\mid 0(\b) \ket, \label{inT}\\
\mid -q_{N},-q_{N-1}, \cdots, q_{1})^{out}_{\t,\b}~&=&~a^{\dagger out}_{-q_{N},\b} 
a^{\dagger out}_{-q_{N-1},\b}
\cdots a^{\dagger out}_{-q_{1},\b}\mid 0(\b) \ket
\label{outT}
\eeq
The S-matrix element is the scalar product of in- and out-states. Its connected part  reads:
\beq
\cS^{\t,\b}(-q_N,-q_{N-1},\cdots ,-q_1;&&q_{N+M},q_{N+M-1},\cdots, q_{N+1})\\
=~~~\int &&\prod^{N+M}_{i=1} d^4x_i~e^{-iq_i\cdot x_i}~i(\partial_i^2+m^2) 
G_{N+M}^{\theta,\b}(x_1,x_2,\cdots,x_{M+N}) 
\label{ncLSZT}
\eeq
where $G_{N+M}^{\theta,\b}$ is:
\begin{eqnarray}
G_{N+M}^{\theta,\b} (x_1, \cdots x_{N+M}) &=& T e^{\frac{i}{2}\sum_{I<J}\partial_{x_I}
\wedge \partial_{x_J}} W_{N+M}^{0,\b} (x_1, \cdots x_{N+M}) \nonumber \\
:&=&~T~W_{N+M}^{\theta,\b} (x_1, \cdots x_{N+M})
\label{GNthetaT}
\end{eqnarray}
and $W_{N+M}^{0,\b}$ are the standard Wightman functions for 
untwisted fields at finite temperature:
\begin{equation}
W_{N+M}^{0,\b} (x_1, \cdots x_{N+M}) = ( 0(\b) | \Phi_0 (x_1) \cdots \Phi_0(x_{N+M})
|0(\b) ).
\end{equation}
This can be derived as in the standard LSZ formalism. 

Note that already for $T~=~0,~(\b~=~\infty)$, $\cS^{\t\infty}~\equiv \cS^\t$ and $S^\t$
do not agree unless $\t~=~0$. This was pointed out in our earliar paper \cite{baltrg}
and remarked on at the beginning of this section. 

In \cite{baltrg} we have indicated how calculations can be performed using eq.\ref{ncLSZ0}
for $T=0$. They  can be extended to $T \neq 0$.
An actual calculation will be presented in a future work.

{\bf Acknowledgments:} The work of APB is supported in part by US-DOE
under grant number DE-FG02-85ER40231. 
The work of APB and TRG are supported by the DST CP-STIO program. 
TRG thanks Prof Hermann Nicolai for a visit to AEI, MPI, Potsdam
when the paper was completed.

\end{document}